\documentclass[twocolumn,pra]{revtex4}
\usepackage{float}
\usepackage{makeidx}
\usepackage{graphicx}

\begin{document}

\title{Continuous-Variable Telecloning with Phase-Conjugate Inputs}
\author{Jing Zhang$^{\dagger }$, Changde Xie, Kunchi Peng}
\affiliation{State Key Laboratory of Quantum Optics and Quantum
Optics Devices, Institute of Opto-Electronics, Shanxi University,
Taiyuan 030006, P.R.China}
\author{Peter van Loock}
\affiliation{Optical Quantum Information Theory Group,
Institute of Theoretical Physics I and Max-Planck Research Group,
Institute of Optics, Information and Photonics,
Universit\"{a}t Erlangen-N\"{u}rnberg, Staudtstr. 7/B2,
91058 Erlangen, Germany}

\begin{abstract}
We propose a scheme for continuous-variable telecloning with
phase-conjugate inputs (PCI). Two cases of PCI telecloning are
considered. The first case is where PCI telecloning produces $M$
clones nonlocally and $M$ anticlones locally, or vice versa.
This kind of PCI telecloning
requires only one EPR (Einstein-Podolsky-Rosen) entangled, two-mode
squeezed state as a resource for building the appropriate multi-mode,
multipartite entangled state via linear optics.
The other case is a PCI telecloning protocol in which both clones and
anticlones are created nonlocally. Such a scheme requires two
EPR entangled states for the generation of a suitable multipartite
entangled state. As our schemes are reversible,
optimal cloning fidelities are achieved in the limit of infinite squeezing.
\end{abstract}

\maketitle

\section{Introduction}

Arbitrary quantum states cannot be copied perfectly according
to the quantum mechanical no-cloning theorem \cite{WoottersZurek,Dieks}.
However, there are optimal quantum cloning protocols that lead to imperfect,
approximate copies resembling the input state just as much as allowed
by quantum theory. This quantum cloning plays an important role in
quantum information and quantum communication.
For instance, it has been shown that quantum cloning might improve the
performance of some computational tasks \cite{one}.
Quantum cloning also represents a potential eavesdropping attack
in quantum cryptographic protocols
\cite{two}.

An approximate, optimal quantum
cloning machine was first considered by Bu\u{z}ek and Hillery \cite{three}.
Their cloner operates in the domain of discrete variables and acts upon
qubit states. Later, quantum cloning was extended to the continuous-variable
(CV) regime by Cerf et al. \cite{four}. Continuous-variable quantum cloning
has been extensively studied in recent years. This interest has been partly
motivated by the fact that preparing and
manipulating optical, Gaussian, CV quantum states is relatively easy compared
to other implementations. There are various theoretical proposals
for an experimental realization of CV quantum cloning
\cite{five,six,seven}. These first proposals rely upon linear optics, and,
in addition, on the optical amplification of the input states.
The cloned states in these schemes are produced locally, from a number of
identical copies of the signal state. We refer to this kind of quantum cloning
as {\it local, conventional} cloning. A more recent experimental realization of local,
conventional cloning of optical coherent states was achieved without
amplification, using only linear optics, homodyne detection, and feedforward
\cite{twenty-one}. According to another interesting proposal of this kind of
cloning \cite{eight}, approximate copies of an optical quantum state appear
in two atomic ensembles.

There has also been a lot of interest in quantum {\it nonlocal} cloning
(telecloning),
which is a combination of quantum cloning and teleportation to more
than just one receiver. The aim of telecloning is to broadcast
information of an unknown quantum state from a sender to several spatially
separated receivers exploiting multipartite entanglement as a multi-user quantum
channel. For continuous variables, the first proposal for optimal
$1\rightarrow M$ telecloning of coherent states is based upon an
$M+1$-partite entangled multi-mode Gaussian state \cite {nine}.
The protocol itself, similar to one-to-one quantum teleportation
\cite{SamKimble}, uses beam splitters, homodyne detection, and feedforward.
In this case, the anticlones
(phase-conjugate clones, or time-reversed state) are lost; thus,
optimal telecloning can be achieved by exploiting nonmaximum, effectively
bipartite entanglement produced from {\it finitely} squeezed light
via linear optics.
This scheme (see also \cite{Ferraro,Adesso}) is regarded as the CV
{\it irreversible} telecloner, analogous to the irreversible
telecloner in the domain of discrete variables \cite{ten}. Recently,
irreversible telecloning of optical coherent states was demonstrated
experimentally \cite{eleven}.

In addition, CV {\it reversible}
telecloning was proposed \cite{eleven1}, in which the
information of an unknown quantum state is, in principle, transferred without loss
from a sender to several spatially separated receivers, again exploiting multipartite
entanglement as a multi-user quantum channel. However, in this case,
optimal reversible telecloning requires maximum bipartite entanglement; hence
infinite squeezing would be needed to build the corresponding multi-mode entangled
state \cite{nine}.

%
\begin{table}
\centerline{
\includegraphics[width=3.3in]{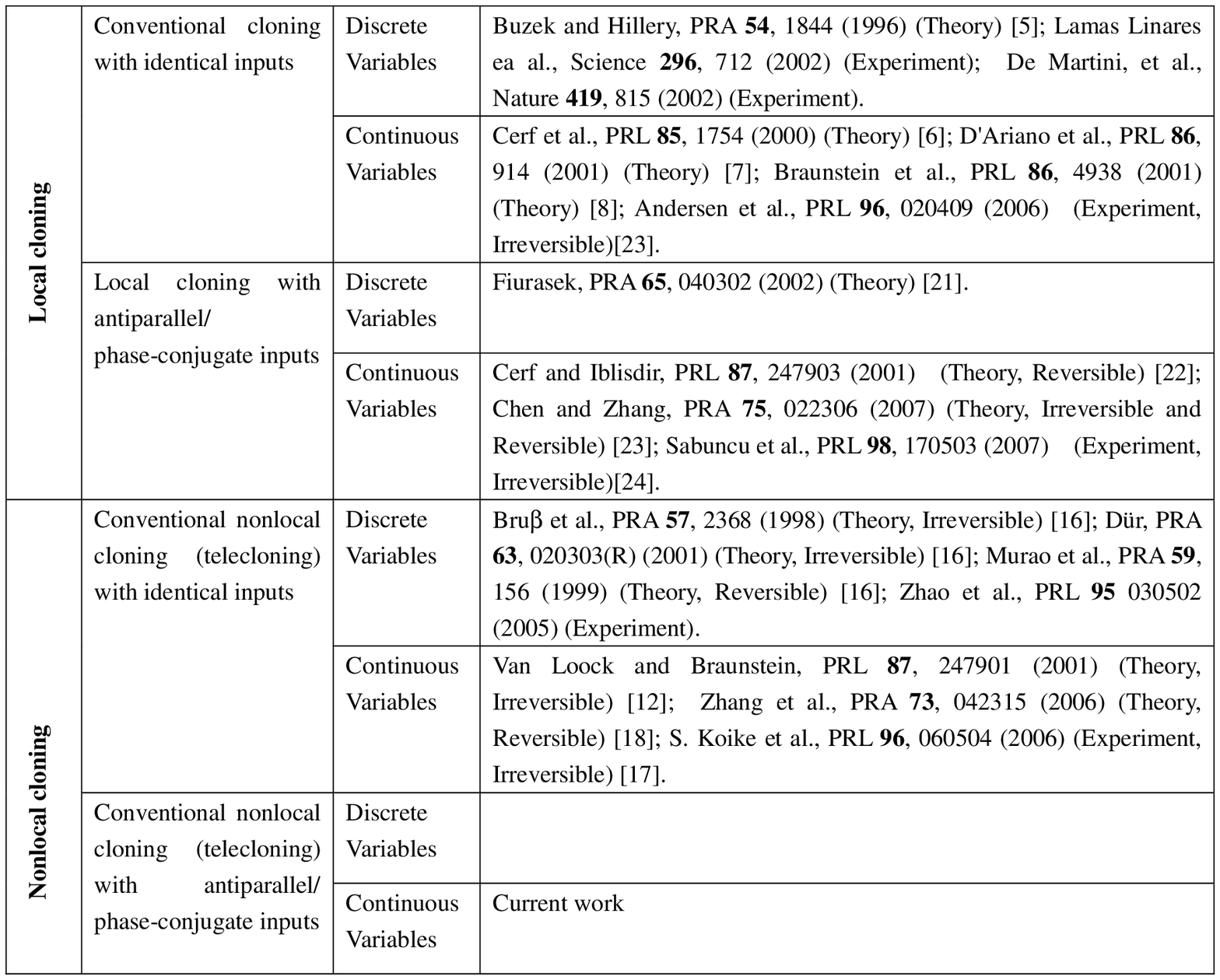}
} \vspace{0.2in}
\caption{Summary of earlier work and existing results
on quantum cloning. \label{Table1} }
\end{table}

An important work on quantum state estimation has revealed that more
quantum information can be encoded in antiparallel pairs of spins
than in parallel pairs \cite{twelve}. Subsequently, a similar result
was obtained in the CV regime, where a pair of conjugate Gaussian
states carries more information than a pair of identical coherent
states \cite{thirteen}. These results enable one to achieve better
fidelities with a cloning machine admitting antiparallel input
qubits or phase-conjugate input coherent states, compared to the
conventional case with identical input copies. Based upon the above
properties, Fiur\'{a}\u{s}ek proposed a cloning machine for
antiparallel spin states \cite {forteen0}. Similarly, Cerf and
Iblisdir derived a CV cloning transformation \cite {forteen} that
uses $N$ copies of a coherent state and $N^{\prime}$ copies of its
complex conjugate as input states, and produces $M$ optimal clones
of the coherent state and $M^{\prime }=M+N^{\prime }-N$
phase-conjugate clones (anticlones, or time reversed states). This
is the first scheme for a local, phase-conjugate input (PCI) cloner
of continuous variables. Nonetheless, an experimental realization of
the proposed PCI cloner is difficult, as it requires ``online''
optical parametric amplification. Recently, a much simpler and
efficient CV PCI cloning machine based on linear optics, homodyne
detection, and feedforward was proposed \cite {fifteen,sixteen} and
implemented experimentally \cite{sixteen}. Note that the production
of an infinite number of clones ($M\rightarrow \infty$) coincides
with optimal or perfect state estimation \cite {estimation}. The
case of $M$ clones and $M-2$ anticlones from two identical replicas
gives the optimal telecloning fidelity of $2/3$ for
$M\rightarrow\infty$ and maximally EPR entangled state. This is
consistent with the standard optimal value of $2/3$ for
$1\rightarrow 2$ cloning of coherent states that was obtained in
Ref. \cite{four}. However, the PCI telecloner yields the better
fidelity of $4/5$ owing to the added information in the
phase-conjugate input state. The result of the increased fidelity of
$4/5$ for coherent states with phase-conjugate input modes indicates
that the added information in the input state must be equivalent to
the $1\rightarrow 2$ cloning of a single mode coherent state with
known phase, where the fidelity is also $4/5$ \cite{ Alexanian}. A
summary of the various quantum cloning schemes and their
realizations is shown in Table 1.

In this paper, we propose a protocol of CV reversible
{\it telecloning} of
coherent states with {\it phase-conjugate} input modes. The
$N+N\rightarrow M+M$ quantum telecloning machine yields $M$
identical clones and $M$ identical anticlones from $N$ copies of a
coherent state and $N$ phase-conjugate copies. Here, we
consider two cases of PCI telecloning. In the first case,
PCI telecloning produces $M$ clones nonlocally and $M$ anticlones
locally, or vice versa. Alternatively, both clones and anticlones
may be created nonlocally through PCI telecloning.
Optimal cloning fidelities of
this PCI telecloning require perfect EPR
(Einstein-Podolsky-Rosen) entanglement, i.e., infinite two-mode
squeezing as a resource. However, similar to the conventional (irreversible)
CV telecloning scheme \cite {nine}, no ``online'' optical parametric
amplification is needed.
As shown in Table 1, PCI telecloning of
qubit states has not been investigated yet. Hence our protocol
represents a nice example, where a CV quantum information protocol
is proposed before its qubit counterpart.

\section{PCI Telecloning with nonlocal clones and local anticlones, $1+1\rightarrow M+M$}

The quantum states we consider in this paper are described with the
electromagnetic field annihilation operator $\hat{a}=(\hat{X}+i\hat{P})/2$,
which is expressed in terms of the amplitude $\hat{X}$ and phase $\hat{P}$
quadrature with the canonical commutation relation $[\hat{X},\hat{P}]=2i$.
Without loss of generality, the quadrature operators can be expressed in
terms of a steady state and fluctuating component as $\hat{A}=\langle \hat{A}%
\rangle +\Delta \hat{A}$, with variances of $V_A=\langle \Delta \hat{A}%
^2\rangle $ ($\hat{A}=\hat{X}$ or $\hat{P})$. The input coherent
state and its phase-conjugate state to be cloned will be described
by $\left| \alpha _{in}\right\rangle =\left| \frac 12\left(
x_{in}+ip_{in}\right) \right\rangle $ and $\left| \alpha
_{in}^{*}\right\rangle =\left| \frac 12\left( x_{in}-ip_{in}\right)
\right\rangle $ respectively, where $x_{in}$
and $p_{in}$ are the expectation values of $\hat{X}_{in}$ and $\hat{P}_{in}$.
The cloning machine generates many clones of the input state
characterized by the density operator $\hat{\rho}_{clone}$ and the
expectation values $x_{clone}$ and $p_{clone}$. The quality of the cloning
machine can be quantified by the fidelity, which is the overlap
between the input state and the output state. It is defined by
\cite{seventeen}

\begin{eqnarray}
F &=&\left\langle \alpha _{in}\right| \hat{\rho}_{clone}\left|
\alpha _{in}\right\rangle =\frac 2{\sqrt{( 1+\Delta
^2\hat{X}_{clone})
( 1+\Delta ^2\hat{P}_{clone}) }}  \nonumber \\
&&*\exp \left[ -\frac{( x_{clone}-x_{in}) ^2}{2(1+\Delta ^2%
\hat{X}_{clone}) }-\frac{( p_{clone}-p_{in}) ^2}{2( 1+\Delta
^2\hat{P}_{clone}) }\right].  \label{1}
\end{eqnarray}
In the case of unity gains, i.e., $x_{clone}=x_{in}$, the fidelity is
strongly peaked and becomes

\begin{equation}
F=\frac 2{\sqrt{( 1+\Delta ^2\hat{X}_{clone}) ( 1+\Delta ^2%
\hat{P}_{clone}) }}.  \label{2}
\end{equation}

%
\begin{figure}
\centerline{
\includegraphics[width=3.3in]{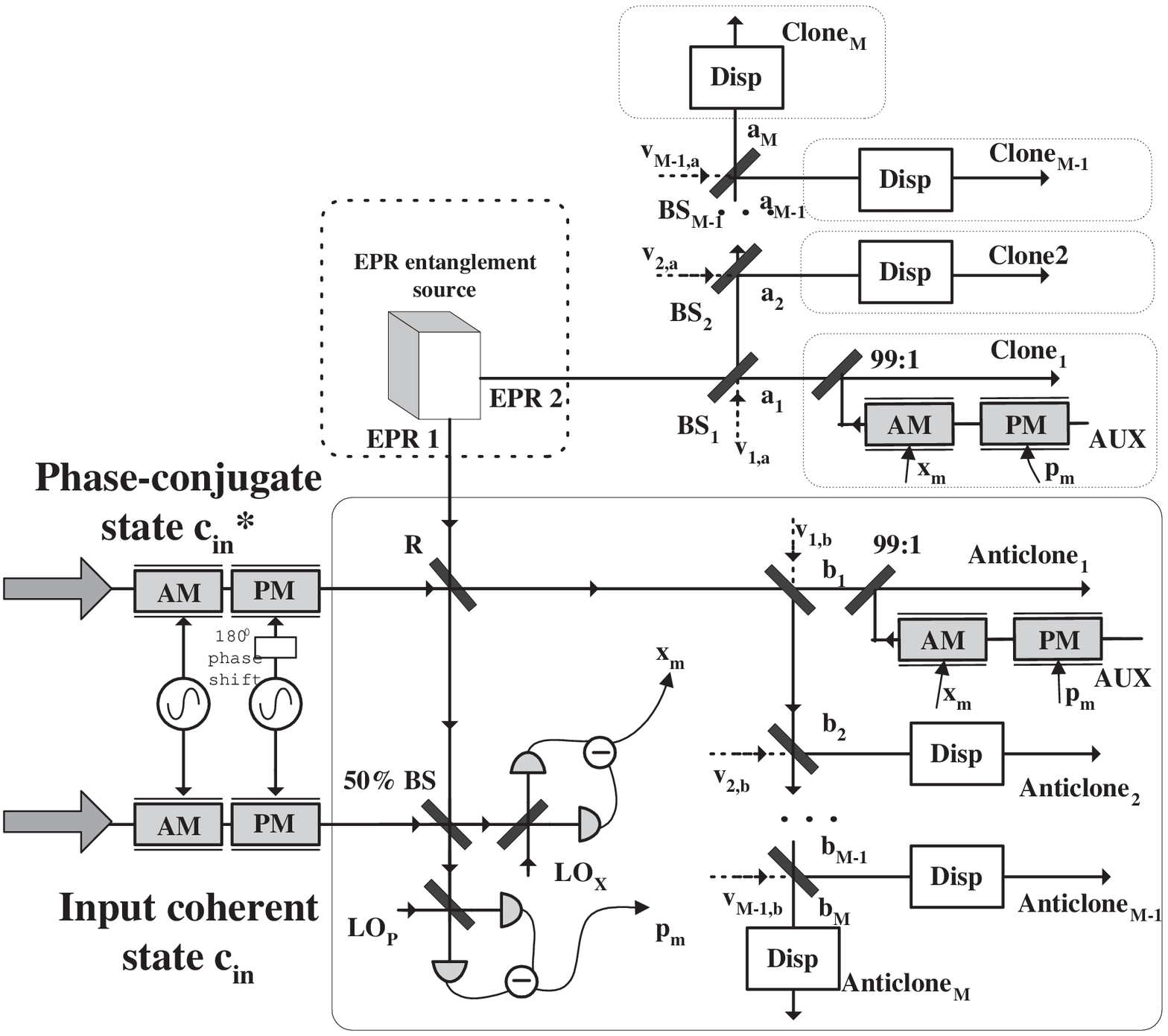}
} \vspace{0.2in}
\caption{ Schematic diagram of $1+1\rightarrow M+M$ PCI telecloning
with nonlocal clones and local anticlones. BS: Beam splitter, LO:
Local oscillator, AM: Amplitude modulator, PM: Phase modulator and
AUX: Auxiliary beam. \label{Fig1} }
\end{figure}

The essence of quantum telecloning is the multipartite entanglement
shared among the sender and the receivers. Without multipartite
entanglement, it is only possible to perform the corresponding
two-step protocol: the sender produces clones and anticlones
locally, and then (bipartitely) teleports them to each receiver. The
two-step protocol would require $2M-1$ bipartite entanglement for
teleportation. Continuous-variable PCI telecloning with nonlocal
clones and local anticlones only requires
one resource of bipartite entanglement.
The bipartite entangled state of CV is a two-mode Gaussian entangled
state (EPR entangled state), which can be obtained directly by
type-II parametric interaction \cite{ninteen} or indirectly by
mixing two independent squeezed beams on a beam-splitter
\cite{seventeen}. The EPR entangled beams have a very strong
correlation property such that both their
sum-amplitude quadrature variance $\langle \delta (\hat{X}%
_{a_{EPR1}}+\hat{X}_{a_{EPR2}})^2\rangle =2e^{-2r}$ and their
difference-phase quadrature variance $\langle \delta (\hat{Y}_{a_{EPR1}}-%
\hat{Y}_{a_{EPR2}})^2\rangle =2e^{-2r}$ are below the two-mode vacuum
noise limit, where $r$ is the squeezing parameter. Let us first
illustrate the protocol in the simple case of $1+1\rightarrow M+M$
PCI telecloning, as shown in Fig. 1. One of the EPR entangled
beams $\hat{a}_{EPR1}$ is hold by the sender and the other
$\hat{a}_{EPR2}$ is distributed among M remote parties via $(M-1)$ beam
splitters with appropriately adjusted transmittances
and reflectivities. The modes $\hat{v}%
_{j,a}$ are in the vacuum state. The EPR
entangled mode $\hat{a}_{EPR2}$ is mixed with $\hat{v}%
_{1,in}$ at the beam splitter $BS_1$. The mode $\hat{a}%
_{1}$ contains the EPR entangled mode $\hat{a}_{EPR2}$ up to a factor
of $1/\sqrt{M}$. The output $\hat{c}_2$ is split at $BS_2$ and
so on, until it arrives at the last beam splitter $BS_{M-1}$.
The transformation performed by the jth beam splitter can be written
as
\begin{eqnarray}
\hat{a}_{j} &=&\sqrt{\frac 1{M-j+1}}\hat{c}_{j,a}+%
\sqrt{\frac{M-j}{M-j+1}}\hat{v}_{j,a},
\label{tele-entanglement} \\
\hat{c}_{j+1,a} &=&\sqrt{\frac{M-j}{M-j+1}}\hat{c}_{j,a}-%
\sqrt{\frac 1{M-j+1}}\hat{v}_{j,a},  \nonumber
\end{eqnarray}
where $\hat{c}_{1,a}=\hat{a}_{EPR2}$, and $%
\hat{a}_{M}=\hat{c}_{M,a}$. It is clearly shown that each mode
$\hat{a}_{j}$ contains a $\sqrt{1/M}$ portion of the EPR entangled
mode $\hat{a}_{EPR2}$ and a $\sqrt{(M-1)/M}$ portion of vacuum.

At the sender station, the input coherent state $\hat{c}_{in}$ and
its phase-conjugate state $\hat{c}_{in}^{*}$ are prepared by an
amplitude modulator and a phase modulator, respectively. The
modulated signals on the amplitude modulators are in-phase and the
modulated signals on the phase modulators are out-of-phase. The
input phase-conjugate state $\hat{c}_{in}^{*}$ is combined with the
EPR entangled beam $\hat{a}_{EPR1}$ via a variable beam splitter
$BS_{0}$ with transmission rate $T$ and reflectivity rate $R$. The
transmitted field
$\hat{c}_{1t}=\sqrt{T}\hat{c}_{in}^{*}-\sqrt{R}\hat{a}_{EPR1}$ is
divided into M modes $\{ \hat{b}_{1},
\hat{b}_{2},...,\hat{b}_{M}\}$ via $(M-1)$ beam splitters, which is
the same as for the EPR entangled beam $\hat{a}_{EPR2}$ (see also Eq.
(\ref{tele-entanglement})). The reflected output
$\hat{c}_{1r}=\sqrt{R}\hat{c}_{in}^{*}+\sqrt{T}\hat{a}_{EPR1}$
is combined with input mode $\hat{%
c}_{in}$ at a 50/50 beam splitter. Then we perform homodyne
measurements on the two output beams in order to obtain the amplitude and
phase quadratures simultaneously. The measured quadratures are

\begin{eqnarray}
\hat{X}_m &=&\frac 1{\sqrt{2}}( \sqrt{R}\hat{X}_{c_{in}^{*}}+\sqrt{T}\hat{X}%
_{EPR1}+\hat{X}_{c_{in}})  \nonumber \\
\hat{P}_m &=&\frac 1{\sqrt{2}}( \sqrt{R}\hat{P}_{c_{in}^{*}}+\sqrt{T}\hat{P}%
_{EPR1}-\hat{P}_{c_{in}}) .  \label{3}
\end{eqnarray}
The sender then conveys the measured results $x_m$ and $p_m$ to the
local modes $\hat{b}_{j}$ and the remote ones $\hat{a}_{j}$. After
receiving the measurement results, each receiver displaces his
corresponding mode by means of a 1/99 beam splitter with an
auxiliary beam, the amplitude and phase of which have been modulated
via two independent modulators using the received $x$ and $p$
signals with the scaling factors $g_{x}=-g_{p}=g_{1}$ for modes
$\hat{a}_{j}$, and, $g_{x}=g_{p}=g_{2}$ for modes $\hat{b}_{j}$
respectively \cite{sixteen}.
Corresponding to the transformation $\hat{A}\rightarrow \hat{D}^{\dagger}\hat{A}%
\hat{D}=\hat{A}+\left( \hat{X}_m+i\hat{P}_m\right) /2$ in the
Heisenberg representation, the displaced fields of the remote
parties can be expressed as

\begin{eqnarray}
\hat{a}_{j}^{^{\prime }} &=&\sqrt{\frac 1{M-j+1}}\hat{c}_{j,a}^{^{\prime }}+%
\sqrt{\frac{M-j}{M-j+1}}\hat{v}_{j,a},
\label{clone} \\
\hat{c}_{j+1,a}^{^{\prime }} &=&\sqrt{\frac{M-j}{M-j+1}}\hat{c}_{j,a}^{^{\prime }}-%
\sqrt{\frac 1{M-j+1}}\hat{v}_{j,a},  \nonumber
\end{eqnarray}
where
\begin{eqnarray}
\hat{c}_{1,a}^{^{\prime }} &=&g_{1}\sqrt{\frac M{2}}\sqrt{R}\hat{c}_{in}^{*\dagger}+g_{1}\sqrt{\frac M{2}}\sqrt{%
1-R}\hat{a}_{EPR1}^{\dagger}\nonumber \\
&&+\hat{a}_{EPR2}+g_{1}\sqrt{\frac M{2}}\hat{c}_{in}  \label{dc} \\
\hat{a}_{M}^{^{\prime }}&=&\hat{c}_{M,a}^{^{\prime }}.
\end{eqnarray}
The displaced local modes can be expressed as

\begin{eqnarray}
\hat{b}_{j}^{^{\prime }} &=&\sqrt{\frac 1{M-j+1}}\hat{c}_{j,b}^{^{\prime }}+%
\sqrt{\frac{M-j}{M-j+1}}\hat{v}_{j,b},
\label{anticlone} \\
\hat{c}_{j+1,b}^{^{\prime }} &=&\sqrt{\frac{M-j}{M-j+1}}\hat{c}_{j,b}^{^{\prime }}-%
\sqrt{\frac 1{M-j+1}}\hat{v}_{j,b},  \nonumber
\end{eqnarray}
where

\begin{eqnarray}
\hat{c}_{1,b}^{^{\prime }} &=&( \sqrt{1-R}+g_{2}\sqrt{\frac M{2}}
\sqrt{R})
\hat{c}_{in}^{*}  \nonumber \\
&&-( \sqrt{R}-g_{2}\sqrt{\frac M{2}}\sqrt{1-R})
\hat{a}_{EPR1}+g_{2}\sqrt{\frac M{2}}\hat{c}_{in}^{\dagger}
\label{5}
\, \\
\hat{b}_{M}^{^{\prime }}&=&\hat{c}_{M,b}^{^{\prime }}.
\end{eqnarray}
By choosing $g_{1}=\sqrt{2/M( 1-R) }$ and $g_{2} =\sqrt{2R/M( 1-R)
}$, the displaced fields $\hat{c}_{1,a}^{^{\prime }}$ and
$\hat{c}_{1,b}^{^{\prime }}$ are given by

\begin{eqnarray}
\hat{c}_{1,a}^{^{\prime }}&=&\frac{\sqrt{R}}{\sqrt{1-R}}\hat{c}_{in}^{*\dagger}+\frac 1{\sqrt{1-R}}%
\hat{c}_{in}
\nonumber \\
&&+(\hat{a}_{EPR1}^{\dagger}+\hat{a}_{EPR2})  \label{disp1}\\
\hat{c}_{1,b}^{^{\prime }}&=&\frac 1{\sqrt{1-R}}\hat{c}_{in}^{*}+\frac{\sqrt{R}}{\sqrt{1-R}}%
\hat{c}_{in}^{\dagger} \label{dispp2}.
\end{eqnarray}
We can see that Eqs. (\ref{disp1}) and (\ref{dispp2}) include a
phase-insensitive amplification with gain $G=1/(1-R)$.

Note that both terms $%
\hat{c}_{in}$ and $\hat{c}_{in}^{*\dagger}$ in Eq. (\ref{disp1})
contribute to the total coherent signal with a factor of
$1/\sqrt{1-R}+\sqrt{R}/\sqrt{1-R}$ and noise variances with
$\sqrt{1+R}/\sqrt{1-R}$, and in Eq. (\ref{dispp2}) they contribute to
the total conjugate coherent signal with a factor of
$1/\sqrt{1-R}+\sqrt{R}/\sqrt{1-R}$ and noise variances with
$\sqrt{1+R}/\sqrt{1-R}$. Since each output cloner
$\hat{a}_{j}^{^{\prime }}$ and anticlone $\hat{b}_{j}^{^{\prime }}$
should include one part of the input coherent and the conjugate state,
$R$ must satisfy

\begin{equation}
\frac 1{\sqrt{M(1-R)}}+\frac{\sqrt{R}}{\sqrt{M(1-R)}}=1. \label{8}
\end{equation}
The parameter $R$ can be easily determined by solving the above equation,

\begin{equation}
R=\frac{\left( M-1\right) ^2}{\left( M+1\right) ^2}.  \label{9}
\end{equation}
According to Eqs. (\ref{clone},\ref{anticlone},\ref{9}), the
variances of the clones and anticlones can be written as

\begin{eqnarray}
\Delta ^2\hat{X}_{a_j^{^{\prime }}}=\Delta ^2\hat{P}_{a_j^{^{\prime
}}}&=&1+\frac{( M-1) ^2}{2M^2}+\frac{2e^{-2r}}M, \nonumber \\
\Delta ^2\hat{X}_{b_j^{^{\prime }}} =\Delta ^2\hat{P}_{b_j^{^{\prime
}}}&=&\frac 1M\frac{1+R}{1-R}+\frac{M-1}M  \label{10} \\
&=&1+\frac{\left( M-1\right) ^2}{2M^2}.  \nonumber
\end{eqnarray}
The fidelity can be obtained via Eq. (\ref{2})

\begin{eqnarray}
F_{\left( _1^1\right) \rightarrow \left(
_M^M\right)}^{clone}&=&\frac{4M^2}{4M^2+\left( M-1\right) ^2+4Me^{-2r}},\nonumber \\
F_{\left( _1^1\right) \rightarrow \left(
_M^M\right)}^{anti}&=&\frac{4M^2}{4M^2+\left( M-1\right) ^2}.
\label{11}
\end{eqnarray}
This scheme produces $M$ anticlones locally and $M$ clones
nonlocally. The fidelity of the anticlones is optimal and
independent of the entanglement. However, the fidelity of the
clones does depend on the entanglement. An optimal fidelity
of the clones requires a maximally EPR
entangled state, $r\rightarrow \infty $. Now we compare $M$ clones
and $M$ anticlones from the phase-conjugate input modes with $M$
clones and $M-2$ anticlones from the two identical replicas. The
fidelity of the standard 2-to-M+(M-2) telecloning is given by
\cite{eleven1}

\begin{eqnarray}
F_{2\rightarrow M+(M-2)}^{clone}&=&\frac{2M}{3M-2+2e^{-2r}}  \label{12} \\
F_{2\rightarrow M+(M-2)}^{anti}&=&\frac{2}{3}
\end{eqnarray}
In the special case $M=2$, the standard telecloner can produce
clones perfectly with fidelity equal to one $(r\rightarrow \infty )$
and no anticlones, while the PCI telecloner yields two clones and
two anticlones with fidelity equal to $16/17$ $(r\rightarrow \infty
)$. Obviously, the PCI telecloner yields a better
fidelity than the standard cloning when $M\geq 3$. In the limit of
large $M\rightarrow \infty $, we see $F_{\left( _1^1\right)
\rightarrow \infty }^{clone}=F_{\left( _1^1\right) \rightarrow
\infty }^{anti}=\frac 45$ compared with the standard telecloning
$F_{2\rightarrow \infty }^{clone}=F_{2\rightarrow \infty
}^{anti}=\frac 23$ $(r\rightarrow \infty )$. This shows that more
information can be encoded into a pair of conjugate coherent states
than by using two identical states, which was shown in Refs.
\cite {thirteen,sixteen}. This scheme can be easily modified
in order to realize PCI
telecloning with local clones and nonlocal anticlones; simply
the inputs of the coherent state and its phase-conjugate state
must be swapped.

\section{PCI Telecloning with nonlocal clones and local anticlones, $N+N\rightarrow M+M$}
We now generalize the $1+1\rightarrow M+M$ case to $N+N\rightarrow M+M$
PCI quantum telecloning, which produces $M$ clones nonlocally and
$M$ anticlones locally from $N$ copies of a coherent state
and $N$ copies of its complex conjugate, as illustrated in
Fig. 2. First, we concentrate $N$ identically prepared coherent states $%
\left| \Phi \right\rangle $ described by $\left\{ \hat{a}_{in,l}\right\} $ $%
(l=1,\cdots N)$ into a single spatial mode $\hat{c}_1$ with amplitude $\sqrt{%
N}\Phi $. This operation can be performed by interfering $N$ input modes in $%
N-1$ beam splitters, which yields the mode

%
\begin{figure}
\centerline{
\includegraphics[width=3.3in]{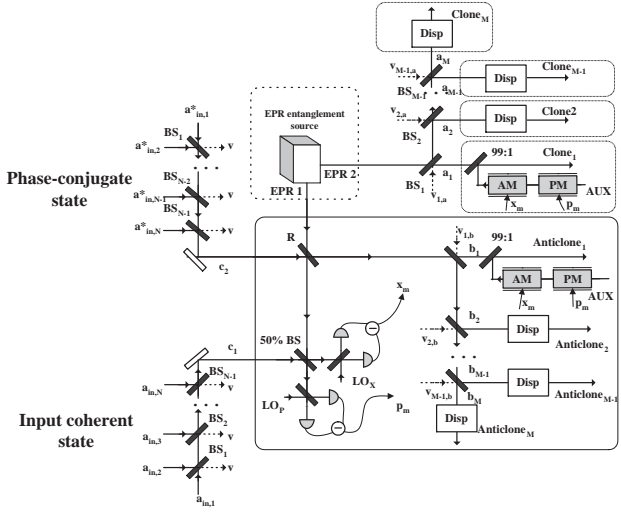}
} \vspace{0.1in}
\caption{ Schematic diagram of $N+N\rightarrow M+M$ PCI telecloning
with nonlocal clones and local anticlones. \label{Fig2} }
\end{figure}

\begin{equation}
\hat{c}_1=\frac 1{\sqrt{N}}\sum_{l=1}^N\hat{a}_{in,l}  \label{13}
\end{equation}
and $N-1$ vacuum modes. The same method can be used for the
generation of the phase-conjugate input mode $\hat{c}_2$ with
amplitude $\sqrt{N}\Phi ^{*}$ from the $N\ $copies of $\left| \Phi
^{*}\right\rangle $ stored in the $N\ $modes $\{ a_{in,l}^{*}\} $
$(l=1,\cdots N)$, which is expressed as
\begin{equation}
\hat{c}_2=\frac 1{\sqrt{N}}\sum_{l=1}^N\hat{a}_{in,l}^{*}.
\label{14}
\end{equation}
Then, $\hat{c}_1$ and $\hat{c}_2$ are sent into the cloning
machine (see Fig. 1). The displaced fields (Eqs. (\ref{disp1}) and
(\ref{dispp2})) become

\begin{eqnarray}
\hat{c}_{1,a}^{^{\prime }}&=&\frac{\sqrt{R}}{\sqrt{1-R}}\hat{c}_{2}^{\dagger}
+\frac 1{\sqrt{1-R}}%
\hat{c}_{1}\nonumber \\
&&+(\hat{a}_{EPR1}^{\dagger}+\hat{a}_{EPR2})  \label{dispN1}\\
\hat{c}_{1,b}^{^{\prime }}&=&\frac 1{\sqrt{1-R}}\hat{c}_{2}+\frac{\sqrt{R}}{\sqrt{1-R}}%
\hat{c}_{1}^{\dagger} \label{dispN2}.
\end{eqnarray}
The terms with $\hat{c}_1$ and $\hat{c}_2^{\dagger}$ in Eqs. (\ref{dispN1})
and (\ref{dispN2}) contribute
to the total coherent signal with a factor of $\sqrt{N}(1/\sqrt{1-R}+\sqrt{R}%
/\sqrt{1-R})$ and noise variances with $\sqrt{(1+R)/(1-R)}$. Since
each output cloner $\hat{a}_{j}^{^{\prime }}$ and anticlone
$\hat{b}_{j}^{^{\prime }}$ should again include one part of the input
coherent and conjugate state, $R$ must satisfy

\begin{equation}
\sqrt{N}(\frac 1{\sqrt{M(1-R)}}+\frac{\sqrt{R}}{\sqrt{M(1-R)}})=1.
\end{equation}
The parameter $R$ can be easily determined by solving the above equation,

\begin{equation}
R=\frac{\left( M-N\right) ^2}{\left( M+N\right) ^2}.
\end{equation}
The variance and fidelity of the $\left( _N^N\right) \rightarrow $
$\left( _M^M\right)$\ PCI telecloner will is now given by

\begin{eqnarray}
\Delta ^2\hat{X}_{a_j^{^{\prime }}}=\Delta ^2\hat{P}_{a_j^{^{\prime
}}}&=&1+\frac{( M-N) ^2}{2M^2N}+\frac{2e^{-2r}}M, \nonumber \\
\Delta ^2\hat{X}_{b_j^{^{\prime }}} =\Delta ^2\hat{P}_{b_j^{^{\prime
}}} &=&1+\frac{\left( M-N\right) ^2}{2M^2N},  \nonumber\\
F_{\left( _N^N\right) \rightarrow \left(
_M^M\right)}^{clone}&=&\frac{4M^2N}{4M^2N+( M-N) ^2+4MNe^{-2r}},\nonumber\\
F_{\left( _N^N\right) \rightarrow \left(
_M^M\right)}^{anti}&=&\frac{4M^2N}{4M^2N+( M-N) ^2}.\label{FN2}
\end{eqnarray}
Obviously, Eq. (\ref{11}) can be obtained from Eq. (\ref{FN2}) with
$N=1$. This result also coincides with that obtained in Ref.
\cite{forteen}. However, the output anticlones are lost in that
scheme. The advantage of dealing with $N\ $pairs of complex conjugate
inputs can be most easily illustrated in the limit of an infinite
number of clones, $M\rightarrow \infty $; from Eq. (\ref{FN2}) we
obtain $F_{_{\left( _N^N\right) \rightarrow \left(
_M^M\right)}}^{clone}=\frac{4N}{4N+1}$ and $F_{_{\left( _N^N\right)
\rightarrow \left( _M^M\right)}}^{anti}=\frac{4N}{4N+1}$, while the
standard telecloning fidelities are $F_{2N\rightarrow
M+(M-2N)}^{clone}=\frac{2N}{2N+1}$ and $F_{2N\rightarrow
M+(M-2N)}^{anti}=\frac{2N}{2N+1}$ $(r\rightarrow \infty )$.

\section{PCI Telecloning with both clones and anticlones nonlocal}

We now consider $1+1\rightarrow M+M$ PCI telecloning, which nonlocally
produces at the same time  $M$ clones and $M$ anticlones from a
coherent state and its phase conjugate using multipartite
entanglement, as shown in Fig. 3. The case of $N+N\rightarrow M+M$ is
easily obtained from the case of
$1+1\rightarrow M+M$ in a similar way to the discussion
of the preceding sections.

%
\begin{figure}
\centerline{
\includegraphics[width=3.3in]{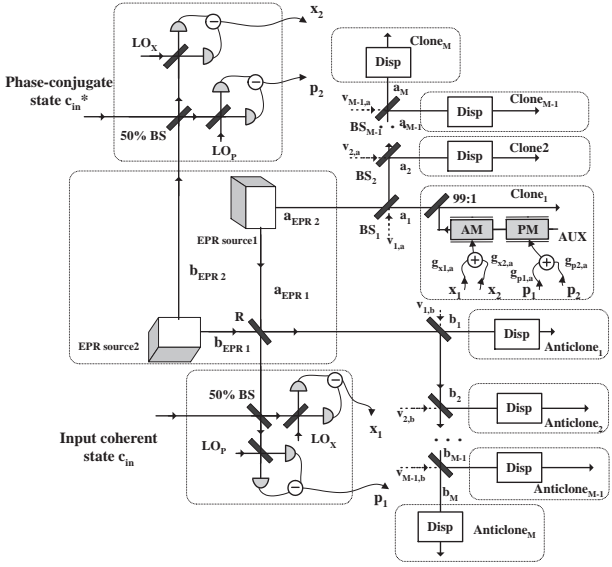}
} \vspace{0.1in}
\caption{ Schematic diagram of $1+1\rightarrow M+M$ PCI telecloning
with both clones and anticlones nonlocal. \label{Fig3} }
\end{figure}

Here, two pairs
$(\hat{a}_{EPR1},\hat{a}_{EPR2})$ and
$(\hat{b}_{EPR1},\hat{b}_{EPR2})$ of EPR entanglement are utilized
with squeezing $r_1$ and $r_2$, respectively. One of the
EPR entangled beams $\hat{a}_{EPR2}$ is distributed among $M$ remote
parties $\{\hat{a}_{1},\hat{a}_{2},...,\hat{a}_{M}\}$ via $(M-1)$
beam splitters similar to Eq. (\ref{tele-entanglement}). One of the
other EPR entangled beams $\hat{b}_{EPR1}$ is combined with the EPR
entangled beam $\hat{a}_{EPR1}$ at a beam splitter $BS_{0}$ with
transmission rate $T=1-R$ and reflectivity $R=\left( M-1\right)
^2/\left( M+1\right) ^2$. The transmitted field
$\hat{c}_{1t}=\sqrt{T}\hat{b}_{EPR1}-\sqrt{R}\hat{a}_{EPR1}$ is also
divided into M remote modes
$\{\hat{b}_{1},\hat{b}_{2},...,\hat{b}_{M}\}$ by $(M-1)$ beam
splitters, similar to the EPR entangled beam $\hat{a}_{EPR2}$.
The reflected output
$\hat{c}_{1r}=\sqrt{R}\hat{b}_{EPR1}+\sqrt{T}\hat{a}_{EPR1}$ and
$\hat{b}_{EPR2}$ are held by the senders 1 and 2, respectively.
Sender 1 combines the reflected
output mode $\hat{c}_{1r}$ with input mode $\hat{%
c}_{in}$ at a 50/50 beam splitter and sender 2 does the same thing
with the EPR entangled beam $\hat{b}_{EPR2}$ and the phase-conjugate
input mode. Note that the modes $\hat{a}_{EPR2}$, $\hat{c}_{1t}$,
$\hat{c}_{1r}$ and $\hat{b}_{EPR2}$ form a genuine four-mode
entangled state, whose properties are different from CV
Greenberger-Horne-Zeilinger (GHZ) and cluster states and discussed
in Ref. \cite{twenty-two}. Next the senders perform homodyne
measurements on the two output beams of the 50/50 beam splitters to
obtain two amplitude and phase-quadrature measurement results
$(x_1,p_1),(x_2,p_2)$ to be conveyed to the remote parties. The
measured quadratures are

\begin{eqnarray}
\hat{X}_1 &=&\frac 1{\sqrt{2}}( \sqrt{R}\hat{X}_{b_{EPR1}}+\sqrt{T}\hat{X}%
_{a_{EPR1}}+\hat{X}_{c_{in}}),  \nonumber \\
\hat{P}_1 &=&\frac 1{\sqrt{2}}( \sqrt{R}\hat{P}_{b_{EPR1}}+\sqrt{T}\hat{P}%
_{a_{EPR1}}-\hat{P}_{c_{in}}), \nonumber \\
\hat{X}_2 &=&\frac 1{\sqrt{2}}( \hat{X}_{b_{EPR2}}+\hat{X}_{c_{in}^{*}}),  \nonumber \\
\hat{P}_2 &=&\frac 1{\sqrt{2}}(
\hat{P}_{b_{EPR2}}-\hat{P}_{c_{in}^{*}}). \label{3}
\end{eqnarray}
After receiving the measurement results, each party in the set
$\{\hat{a}_{1},\hat{a}_{2},...,\hat{a}_{M}\}$ combines these
results,
\begin{eqnarray}
x_{s,a}&=&g_{x1,a}x_1+g_{x2,a}x_2 \nonumber \\
&=&\frac 1{\sqrt{M}}( \sqrt{\frac {R}{1-R}}\hat{X}_{c_{in}^{*}}%
+\hat{X}_{a_{EPR1}}+\sqrt{\frac {1}{1-R}}\hat{X}_{c_{in}})\nonumber \\
&&+\sqrt{\frac {R}{M(1-R)}}(\hat{X}_{b_{EPR1}}+\hat{X}_{b_{EPR2}}),\nonumber \\
p_{s,a}&=&g_{p1,a}p_1+g_{p2,a}p_2 \nonumber \\
&=&-\frac 1{\sqrt{M}}( \sqrt{\frac {R}{1-R}}\hat{P}_{c_{in}^{*}}%
+\hat{P}_{a_{EPR1}}-\sqrt{\frac {1}{1-R}}\hat{P}_{c_{in}}) \nonumber \\
&&+\sqrt{\frac {R}{M(1-R)}}(\hat{P}_{b_{EPR2}}-\hat{P}_{b_{EPR1}}),
\end{eqnarray}
where
$g_{x1,a}=-g_{p1,a}=g_{x2,a}/\sqrt{R}=-g_{p2,a}/\sqrt{R}=\sqrt{2/M(1-R)}$,
and finally displaces the corresponding entangled mode. The output fields are the
clones of PCI telecloning with the variances and fidelity given by
\begin{eqnarray}
\Delta ^2\hat{X}_{a_j^{^{\prime }}}&=&\Delta
^2\hat{P}_{a_j^{^{\prime
}}}=1+\frac{( M-1) ^2}{2M^2}(1+e^{-2r_2})+\frac{2e^{-2r_1}}M, \nonumber \\
F_{\left( _1^1\right) \rightarrow \left(
_M^M\right)}^{clone}&=&\frac{4M^2}{4M^2+(
M-1)^2(1+e^{-2r_2})+4Me^{-2r_1}}.\nonumber \\
\end{eqnarray}
Similarly, each party in the set $\{\hat{b}_{1},\hat{b}_{2},...,\hat{b}_{M}\}$
combines the measurement results,
\begin{eqnarray}
x_{s,b}&=&g_{x1,b}x_1+g_{x2,b}x_2 \nonumber \\
&=&\frac 1{\sqrt{M}}( \sqrt{\frac {1}{1-R}}\hat{X}_{c_{in}^{*}}%
+\sqrt{R}\hat{X}_{a_{EPR1}}+\sqrt{\frac {R}{1-R}}\hat{X}_{c_{in}}\nonumber \\
&&+\frac {1}{\sqrt{1-R}}\hat{X}_{b_{EPR2}})+\frac
{R}{\sqrt{M(1-R)}}\hat{X}_{b_{EPR1}},\nonumber \\
p_{s,b}&=&g_{p1,b}p_1+g_{p2,b}p_2 \nonumber \\
&=&\frac 1{\sqrt{M}}( \sqrt{\frac {1}{1-R}}\hat{P}_{c_{in}^{*}}%
+\sqrt{R}\hat{P}_{a_{EPR1}}-\sqrt{\frac {R}{1-R}}\hat{P}_{c_{in}} \nonumber \\
&&-\frac 1{\sqrt{1-R}}\hat{P}_{b_{EPR2}})+\frac
{R}{\sqrt{M(1-R)}}\hat{P}_{b_{EPR1}},
\end{eqnarray}
where $g_{x1,b}=g_{p1,b}=\sqrt{2R/M(1-R)}$ and
$g_{x2,b}=-g_{p2,b}=\sqrt{2/M(1-R)}$,
and finally displaces the corresponding entangled mode.
Now the output fields are the anticlones of PCI telecloning with
the variances and fidelity
\begin{eqnarray}
\Delta ^2\hat{X}_{b_j^{^{\prime }}}&=&\Delta
^2\hat{P}_{b_j^{^{\prime
}}}=1+\frac{( M-1) ^2}{2M^2}+\frac{( M+1) ^2}{2M^2}e^{-2r_2}, \nonumber \\
F_{\left( _1^1\right) \rightarrow \left(
_M^M\right)}^{anti}&=&\frac{4M^2}{4M^2+( M-1)^2+( M+1)^2e^{-2r_2}}.
\end{eqnarray}
This protocol produces nonlocally $M$ clones and $M$ anticlones
at the same time. The fidelity of the clones and anticlones depends on
the entanglement. Clearly, an optimal fidelity
of the clones and anticlones, in agreement with the results
of Ref. \cite{thirteen,ninteen}, requires a perfectly entangled
state, $r_1, r_2\rightarrow \infty $.

\section{Conclusion$\ $}

In conclusion, we have proposed a protocol for CV telecloning of
coherent states with phase-conjugate input (PCI) modes. Two kinds of the
PCI telecloning are considered. In the first scheme,
the PCI telecloning produces
clones nonlocally and anticlones locally. Realization
of this kind of PCI
telecloning requires a single EPR (two-mode squeezed)
entangled state as a resource, arrays of beam splitters,
homodyne detection, and feedforward.
Through the alternative scheme,
both clones and anticlones are produced nonlocally at
the same time. This scheme requires two EPR entangled
states as a resource.
The protocols described here may be applicable to various quantum
communication scenarios, e.g., to an eavesdropping attack in
quantum key distribution.

$^{\dagger} $Corresponding author's email address:
jzhang74@yahoo.com, jzhang74@sxu.edu.cn

\acknowledgments

\textbf{\ }This research was supported in part by NSFC for
Distinguished Young Scholars (Grant No. 10725416), National
Fundamental Research Natural Science Foundation of China (Grant No.
2006CB921101), NSFC (Grant No. 60678029, 60736040), Doctoral Program
Foundation of Ministry of Education China (Grant No. 20050108007),
Program for Changjiang Scholars and Innovative Research Team in
University, Natural Science Foundation of Shanxi Province (Grant No.
2006011003), and the Research Fund for the Returned Abroad Scholars
of Shanxi Province. PvL acknowledges funding from the DFG under the
Emmy Noether programme.


\begin{thebibliography}{99}

\bibitem{WoottersZurek} W.\ K.\ Wootters and W.\ H.\ Zurek, Nature {\bf 299},
802 (1982).

\bibitem{Dieks} D.\ Dieks, Phys. Lett. {\bf 92A}, 271 (1982).

\bibitem{one}  E. F. Galvao and L. Hardy, Phys. Rev. A \textbf{62}, 022301
(2000).

\bibitem{two}  N. Gisin, G. G. Ribordy, W. Tittel, and H. Zbinden,
Rev. Mod. Phys. {\bf 74}, 145 (2002).

\bibitem{three}  V. Bu\u{z}ek and M. Hillery, Phys. Rev. A \textbf{54}, 1844
(1996).

\bibitem{four}  N. J. Cerf, A. Ipe, and \ X. Rottenberg, Phys. Rev. Lett.
\textbf{85}, 1754 (2000).

\bibitem{five}  G. M. D'Ariano, F. De Martini, and M. F. Sacchi, Phys. Rev.
Lett. \textbf{86}, 914 (2001).

\bibitem{six}  S. L. Braunstein, N. J.Cerf, S. Iblisdir, P.van Loock, and S.
Massar, Phys. Rev. Lett. \textbf{86}, 4938 (2001).

\bibitem{seven}  J. Fiurasek, Phys. Rev. Lett. \textbf{86}, 4942 (2001).

\bibitem{twenty-one}  V. Josse, M. Sabuncu, N. J. Cerf, G. Leuchs, and
U. L. Andersen, Phys. Rev. Lett. \textbf{96}, 163602 (2006).

\bibitem{eight}  J. Fiurasek, N. J. Cerf, E. S. Polzik, Phys. Rev. Lett. \textbf{93},
180501 (2004).

\bibitem{nine}  P. van Loock and S.\ L.\ Braunstein, Phys. Rev. Lett.
\textbf{87}, 247901 (2001).

\bibitem{SamKimble} S.\ L.\ Braunstein and H.\ J.\ Kimble,
Phys.\ Rev.\ Lett.\ {\bf 80}, 869 (1998).

\bibitem{Ferraro} A. Ferraro and M. G. A. Paris,
Phys. Rev. A {\bf 72}, 032312 (2005).

\bibitem{Adesso} G. Adesso, A. Serafini, and F. Illuminati,
New J. Phys. {\bf 9}, 60 (2007).

\bibitem{ten}  D. Bru$\beta $ $et$ $al$., Phys. Rev. A \textbf{57}, 2368
(1998); W. D$\ddot{u}$r, Phys. Rev. A \textbf{63}, 020303 (2001); M.
Murao et al., Phys. Rev. \textbf{59}, 156 (1999).

\bibitem{eleven}  S. Koike $et$ $al$., Phys. Rev. Lett.
\textbf{96}, 060504 (2006).

\bibitem{eleven1}  J. Zhang, C. Xie, and K. Peng, Phys. Rev A \textbf{73}, 042315
(2006)

\bibitem{twelve}  N. Gisin and S. Popescu, Phys. Rev. Lett. \textbf{83}, 432 (1999).

\bibitem{thirteen}  N. J. Cerf and S. Iblisdir, Phys. Rev. A \textbf{64}, 032307 (2001).

\bibitem{forteen0}  J. Fiurasek et al., Phys. Rev. A. \textbf{65}, 040302
(2002).

\bibitem{forteen}  N. J. Cerf and S. Iblisdir, Phys. Rev. Lett. \textbf{87}, 247903
(2001).

\bibitem{fifteen}  H. Chen, and J. Zhang, Phys. Rev. A \textbf{75},022306 (2007).

\bibitem{sixteen}  M. Sabuncu, U. L. Andersen, G. Leuchs, Phys. Rev. Lett. \textbf{98},
170503 (2007).

\bibitem{estimation} J. Bae and A. Acin, Phys. Rev. Lett. \textbf{97}, 030402 (2006)

\bibitem{Alexanian} M. Alexanian, Phys. Rev. A $\textbf{73}$, 045801 (2006)

\bibitem{seventeen}  A. Furusawa et al., Science \textbf{282}, 706 (1998).

\bibitem{ninteen}  Z. Y. Ou $et$ $al$., Phys. Rev. Lett. \textbf{68}, 3663
(1992); Y. Zhang $et$ $al$., Phys. Rev. A \textbf{62}, 023813
(2000).

\bibitem{twenty} N. J. Cerf and S. Iblisdir, Phys. Rev. A \textbf{62}, 040301(R) (2000).

\bibitem{twenty-two} J. Zhang, C. Xie, and K. Peng, Phys. Rev A \textbf{76},
064301 (2007)

\end{thebibliography}
\end{document}